\documentclass{aa}  

\usepackage{graphicx}
\usepackage{txfonts}
\usepackage{lipsum}
\usepackage{subcaption}         
\usepackage{lscape}             
\usepackage{placeins}           
\usepackage{makecell}                               

\begin{document}

   \title{Minimum Entropy Indicator for Evaluating Dispersion Measure}

   \subtitle{Maximizing the amount of information}

   \author{Yu-Fu Shen\inst{1}
        \and Yan Xu \inst{1}
        }

   \institute{Changchun Observatory, National Astronomical Observatories, Chinese Academy of Sciences, 130117, Jilin, China\\
             \email{shenyf@cho.ac.cn}
            }

   \date{Received XX XX, 20XX}

 
  \abstract
   {Advanced radio telescopes such as the Five-hundred-meter Aperture Spherical radio Telescope (FAST) can provide high-sensitivity and high-time-resolution data of a large number of radio sources, offering an excellent opportunity for studying radio pulse profiles. However, studying pulse profiles requires the analysis of dispersion measurement (DM). The fitting method tends to make the profile conform to the model, so the fitting method is not suitable for pulse profile research. Indicators are needed to determine the profile closest to the real one, which is the profile discrimination method.}
   {This work is based on the definition of Shannon's information entropy, and believes that the pulse profile when the entropy is minimized is the closest to the true profile. This indicator is simple to calculate and can provide help for pulse profile research.} 
   {This work uses real data from 48 pulsars. By calculating the information entropy of the pulsar profiles under different DMs, the DM corresponding to the minimum entropy is found, thus verifying the validity of the minimum entropy indicator.}
   {In terms of the analysis results for the 48 pulsars, the differences with the references are less than 0.5\% for all except 3 stars, and the results for these 3 stars are consistent with older references. }
   {The minimum entropy indicator can effectively obtain the DMs of radio pulse signals with low computational complexity, but it cannot be proven to be the optimal criterion. It is suggested to use multiple indicators separately when studying pulse profiles. It can be expected that the optimal indicator can provide information on the radiation mechanism of radio sources.}


   \maketitle

\nolinenumbers
\section{Introduction}
Radio sources in the universe are strongly affected by the dispersion measure (DM). The tenuous electron plasma between the source and observer causes a frequency dependent time delay:

\begin{equation}
\Delta t= 1/(2.41 \cdot 10^{-4}) \times\left(v_{\mathrm{low}}^{-2}-v_{\mathrm{high}}^{-2}\right) \times \mathrm{DM}
\end{equation}

where $\Delta t$ is the time delay in seconds, $v_{\mathrm{low}}$ and $v_{\mathrm{high}}$ are the low and high frequencies in MHz, respectively. DM ($\text{pc} \cdot \text{cm}^{-3}$) is defined as

\begin{equation}
\mathrm{DM}=\int_0^d n_{\mathrm{e}} d l
\end{equation}

Thus, when studying the short-term variations of radio sources, it is necessary to determine the DM. One example is the fast radio bursts (FRBs), which are radio flashes that typically last in milliseconds \citep{2007Sci...318..777L}. Most FRBs have larger DM than the maximum estimated DM for the Milky Way \citep{2002astro.ph..7156C,2017ApJ...835...29Y}, so most FRBs are considered extragalactic \citep{2013Sci...341...53T,2019ARA&A..57..417C,2019A&ARv..27....4P,2020Natur.587...45Z}. Another example is pulsars, they are rapidly rotating neutron stars, emit beams of radiation from their magnetic poles, mainly in the radio band. Most pulsars have the spin periods on the order of seconds, but their are also some pulsars known as ``millisecond pulsars'' \citep{1982Natur.300..615B,2013A&A...556A...2V}. To predict the time of arrival (ToA) of the pulse of pulsars, one must consider DM; considering DM also contributes to other effects such as the relationship between DM and ionized media \citep{1990ARA&A..28..561R,1990ApJ...364..123F}. As for the methods to calculate DM, they can be broadly divided into two categories as follows. 

The first category is the fitting method. Timing tools such as tempo\footnote{\url{https://tempo.sourceforge.net/}}, tempo2 \citep{2006MNRAS.369..655H}, and PINT \citep{2021ApJ...911...45L}, use specific template of pulse profile to fit the data, simulating ToA of pulses. Assuming the profile evolves only a little over the observing band, ToA of pulses at different central frequencies can be used to calculate DM, such as Epoch-wise DM \citep[e.g.][base on tempo2]{2019A&A...624A..22D,2019MNRAS.487..394T} and DMX \citep[][based on PINT]{2023ApJ...951L...9A}. The European Pulsar Timing Array\footnote{\url{https://www.epta.eu.org/}} and the Parkes Pulsar Timing Array collaborations \citep{2013PASA...30...17M} use a fully Bayesian-based noise-analysis approach, which could also be classified as the fitting method. The above methods, to some extent, make simplified assumptions about the pulse profile, but in reality, scattering makes the pulse profile widen asymmetrically \citep{2021MNRAS.504.1115O}; the pulse profile evolves with frequency \citep{2019ApJ...871...34P}; their are micropulses \citep{1971ApJ...169..487H,1979AuJPh..32....9C} and sub-pulses that always drift \citep{1968Natur.220..231D,2006A&A...445..243W,2007A&A...469..607W,2019MNRAS.482.3757B}. Attempting to fit the aforementioned phenomena can also yield a DM, but progress in this area is currently limited, primarily due to our insufficient understanding of the pulse emission mechanism. After all, when fitting methods are used, what is obtained is more of a DM that makes the pulsar profile closest to the model rather than the real DM. Thus, the fitting method is not applicable to the study of pulsar profiles, but is effective when studying changes in DM. The second category is more suitable for studying the profile, which will be introduced in the next paragraph.

The second category is to modify the DM until the curve of the total flux within a certain frequency range becomes the most ``real''. This is referred to as the profile discrimination method in this paper. $pdmp$ from $psrchive$ use S/N as an indicator\footnote{\url{http://psrchive.sourceforge.net/manuals/psrstat/algorithms/snr/}}. In the FRB field, they use the Fourier technique to maximize the fine structure (e.g. \citet{2019ascl.soft10004S}\footnote{\url{https://github.com/danielemichilli/DM_phase}} and \citet{2022arXiv220813677L}\footnote{\url{https://github.com/hsiuhsil/DM-power}}). \cite{2022ApJ...935...84L} produces the best DM by making sub-band profiles similar to each other in phase; \cite{2021MNRAS.508.1947T} utilized singular value decomposition (SVD) for the same purpose. \cite{2019ApJ...876L..23H} choose to maximize the difference in profile. The indicators selected by the aforementioned methods can be categorized into three types. Firstly, the S/N indicator might be the first one used. However, the pulse profile with the highest S/N is not necessarily in the finest detail. Additionally, if the contour lacks a distinct peak, the S/N indicator becomes less effective. Secondly, the indicator of consistency among subband profiles may hold when the frequency range is small, but when the frequency range is large, the profiles may inherently be inconsistent. Third, maximizing profile details is a qualitative indicator, and while the mathematical definitions adopted to achieve this goal vary, they are all reasonable and it is difficult to distinguish their merits and demerits. This work continues the idea of maximizing details by introducing the concept of information entropy and directly defining the minimum entropy indicator. Although it cannot be claimed that the minimum entropy profile is definitely more ``real'' than the profiles obtained by other indicators, considering the clarity of the definition and the simplicity of the calculation of the minimum entropy indicator, it is highly recommended as a reference when studying pulse profiles.

\cite{Shannon} proposed information entropy

\begin{equation}
H(X)=-\sum_{i=1}^n p\left(x_i\right) \log p\left(x_i\right)
\end{equation}

where $H$ is the entropy of the random variable $X$, $x_i$ is a possible value of $X$, so in a pulsar, $x_i$ is a short time interval, $p(x_i)$ is the possibility of receiving a unit flux within time interval $x_i$. $n$ is the number of all possible $x_i$. The default base of the logarithm is 2, and in this case, the unit of entropy is bit. Entropy provides a measure of the average amount of information needed to represent an event drawn from a probability distribution for a random variable. We believe, under this definition, within a specific time range, that the flux distribution with the minimum entropy (maximum amount of information) is the closest to the real distribution and corresponds to the optimal DM.

Furthermore, to test the validity of the minimum entropy indicator, this work calculates the DMs of 48 pulsars using this indicator, with data sourced from the Five-hundred-meter Aperture Spherical radio Telescope \citep[FAST][]{2006ScChG..49..129N,2019SCPMA..6259502J,2020RAA....20...64J,2020Innov...100053Q}. FAST is the most sensitive single-dish radio telescope in the world to search for pulsars \citep{2021RAA....21..107H} and to detect hydrogen lines \citep{2022SCPMA..6529702H} or recombination lines \citep{2022SCPMA..6529703H}, offering exceptional opportunities to study the pulse profiles of FRBs and pulsars with high sensitivity and temporal resolution ($4.9152 \cdot 10^{-5} \ \text{s}$). This advancement also places higher demands on the indicators for profile discrimination methods.

\section{Method}\label{method}

\subsection{Data preprocessing}

Regardless of how the data are preprocessed, the entropy of the flux curve can be calculated. The following content on data preprocessing is only specific to the data used in this work and is only for reference. In each slice of the data, we first retain only data points greater than 3 sigma, then divide the data into three segments according to frequency, calculate the sum of data points in each time interval for each segment, and mark the time intervals with sums greater than $3\sigma$. If a time interval is marked in all three segments, it indicates that this is interference occurring at that time interval, and all corresponding data points are assigned the median value of the data. Subsequently, the data is divided into three segments based on time, and the same operation is performed in terms of frequency to remove interference occurring at a certain frequency. Dividing the data into three segments ensures that the true pulse signals are preserved intact, avoiding the removal of any part of the true signals as interference.

\subsection{Pre input parameters}

\begin{figure}[h]
    \centering
    \includegraphics[width=0.95\linewidth]{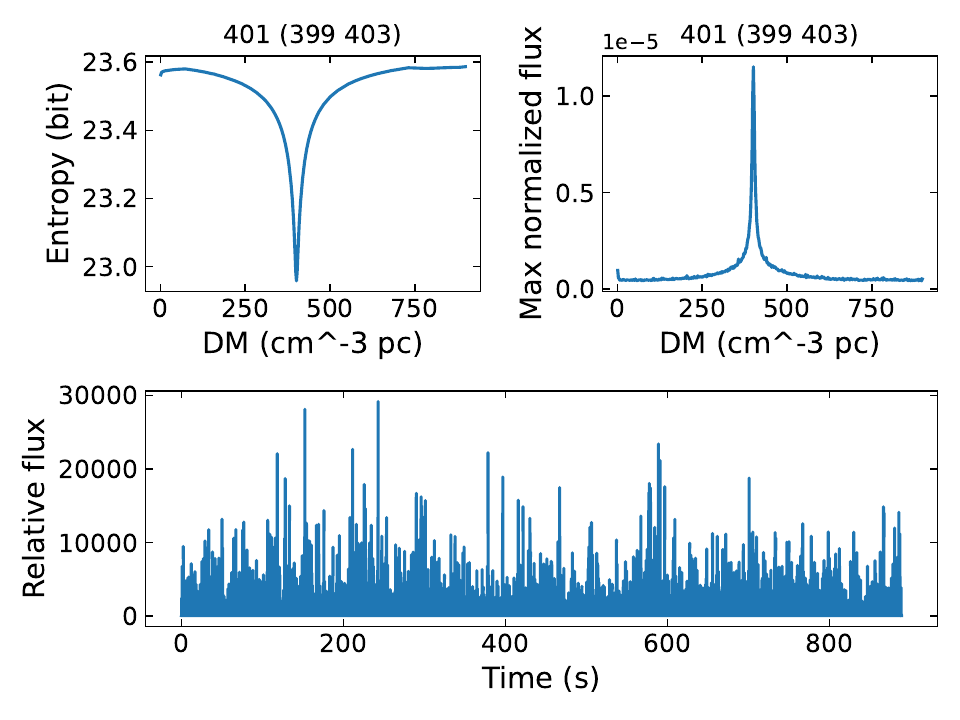}
    \caption{Using pulsar J1901+0331 as an example. Without any pre input, search for DM with an interval of 2 within the range of 1 to 900. The title of the upper two panels are the DMs correspond to the peaks, and the value in parentheses indicates the error range.}
    \label{pre}
\end{figure}

Pre-input the time range of the analyzed data to ensure that the signal is contained within this range. The Minimum Entropy method does not require determining the central position of the signal, as this does not affect the entropy calculation. However, it is still recommended to place the signal near the middle of the set time range to reduce the possibility of missing part of the signal. If the target is a pulsar, by pre-inputting its period and the time when the first signal appears, the DM of each pulse can be conveniently calculated in sequence. If no pre-input is made, directly applying the minimum entropy method to the entire data set (after preprocessing) can still produce a rough reference DM value and help to set the pre input values, as shown in Figure~\ref{pre}. 

\begin{figure}[h]
    \centering
    \includegraphics[width=0.95\linewidth]{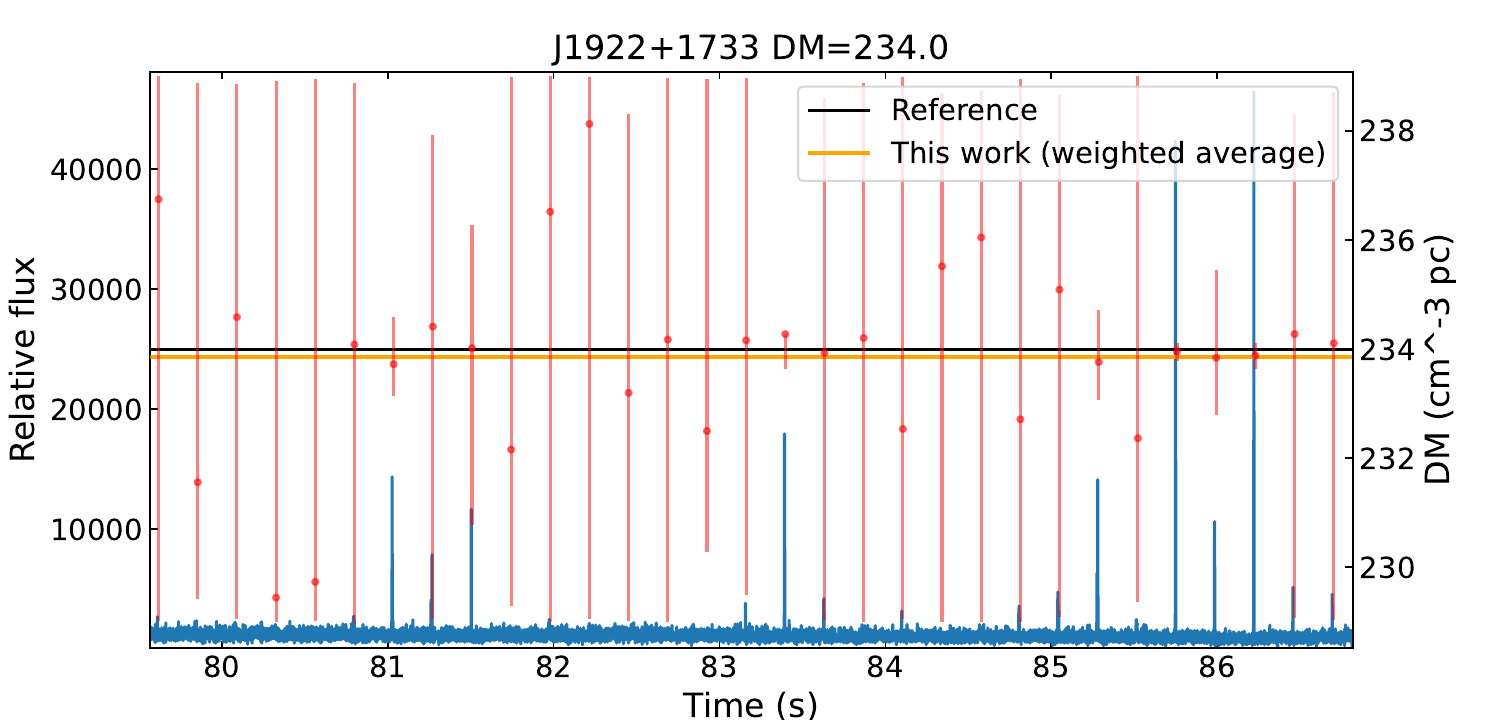}
    \caption{Using pulsar J1901+0331 as an example. Search for DM for each pulse with an interval of 0.001 within the range of 229 to 239. The data points with error bars are DM values given by minimum entropy method. The blue line is the flux curve.}
    \label{showerr}
\end{figure}

We recommend using the following error calculation strategy. After obtaining the curve in the top-left panel of Figure~\ref{pre}, smooth the curve, calculate the difference between them and the standard deviation of this difference, and then determine the DM values corresponding to the range that is greater than the minimum entropy value but does not exceed three times the standard deviation as the error range. After roughly finding the DM, further narrow the DM interval within the error range, and then calculate the DM for each pulse separately. The error range for each pulse using this strategy can reflect the intensity of the pulse, as shown in Figure~\ref{showerr}.

\subsection{Test on synthetic data}

\begin{figure}[h]
    \centering
    \includegraphics[width=0.95\linewidth]{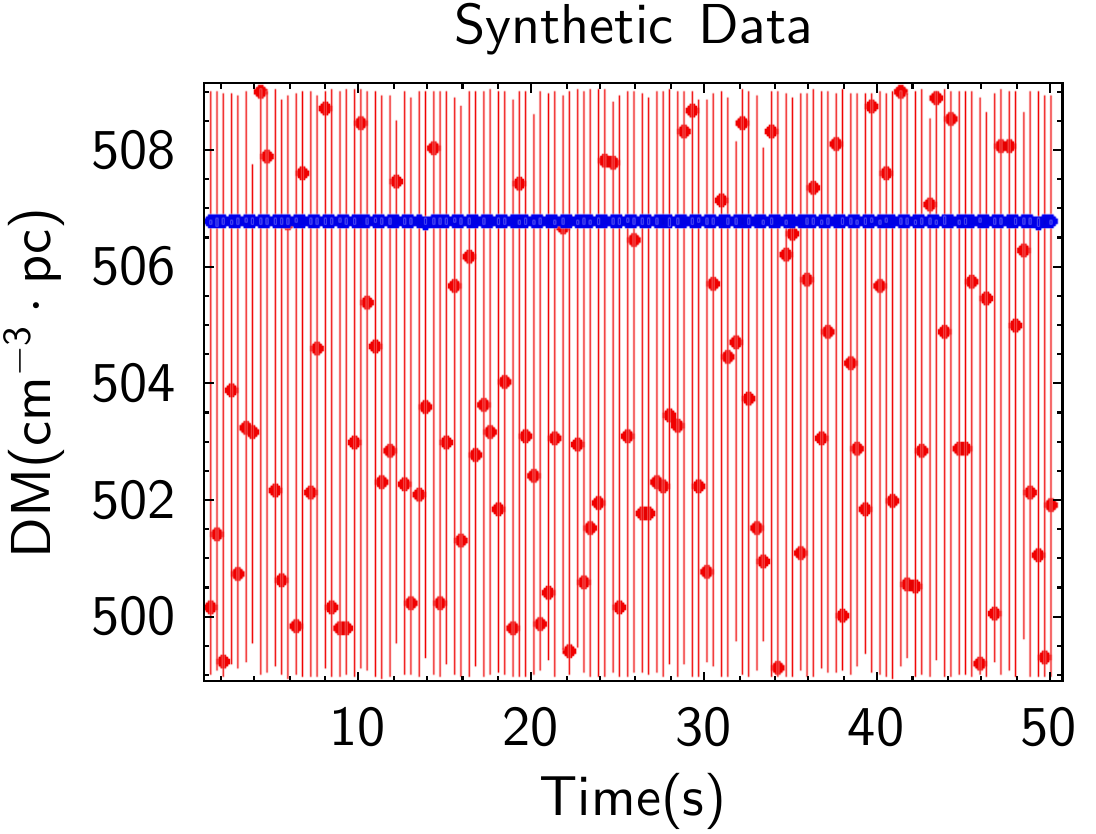}
    \caption{The test conducted using synthetic data. The red points with errorbars are from S/N =0 data. The blue points with errorbars are from S/N = 1.5 data.}
    \label{syn}
\end{figure}

To ensure that the code does not introduce systematic errors, we conducted tests using synthetic pulsar data. The synthetic data consists of a Gaussian noise background combined with Gaussian-distributed signals. The median value of the noise is consistent with the template, and the ratio of the median value of the signal to the median value of the noise is defined as S/N. Both the standard deviations of the signal and the noise are consistent with the template. The flux of the signal is linearly dispersed into the nearest two points based on the distance from the theoretical value to the actual data points. The minimum width unit of the signal corresponds to the time resolution of the data points. Although this method of generating simulated data is rough, it is sufficient to exclude potential systematic errors within the code. The template for the synthetic data is J1857+0212. Figure~\ref{syn} shows the results for S/N of 0 and 1.5, ensuring that the code does not introduce false DM when there is no signal. In the example, the error of DM is approximately 0.007 when S/N is 1.5. After integrating the minimum entropy calculation into other codes, it is recommended to perform similar tests beforehand.

\subsection{Result of 48 pulsars from FAST observations}

The results are shown in Table~\ref{1}. The DMs of each pulse are calculated separately and the ``entropy DMs'' in Table~\ref{1} are weighted averages.

\begin{table*}
\caption{Result DM of 48 pulsars}
\label{1}
\centering   
\begin{tabular}{|l|r|l|r|r|}
\hline
  \multicolumn{1}{|c|}{PSR} &
  \multicolumn{1}{c|}{\makecell{Reference DM \\ ($\text{pc} \cdot \text{cm}^{-3}$)}} &
  \multicolumn{1}{c|}{References} &
  \multicolumn{1}{c|}{\makecell{Entropy DM \\ ($\text{pc} \cdot \text{cm}^{-3}$)}} &
  \multicolumn{1}{c|}{err} \\
\hline
  J0302+2252 & 18.99 & \cite{2016ARep...60..220T,2024ApJS..271...23D} & 19.033 & 0.224\\
  J0628+0909 & 87.9 & \cite{2006ApJ...637..446C,2023RAA....23j4002W} & 88.557 & 0.201\\
  J1404+1159 & 18.4 & \cite{2003PhDT.........2C,2023RAA....23j4002W} & 15.857 & 2.8\\
  J1841+0912 & 49.1 & \cite{1978MNRAS.185..409M,2023RAA....23j4002W} & 49.138 & 0.269\\
  J1842+1332 & 102.5 & \cite{2001MNRAS.326..358E,2023RAA....23j4002W} & 102.331 & 2.403\\
  J1851+1259 & 70.2 & \cite{1985Natur.317..787S,2023RAA....23j4002W} & 70.605 & 0.073\\
  J1851-0053 & 24.0 & \cite{2004MNRAS.352.1439H,2023RAA....23j4002W} & 24.425 & 0.542\\
  J1852+0031 & 787.0 & \cite{Clifton86,2023RAA....23j4002W} & 786.602 & 2.742\\
  J1853+0505 & 260.0 & \cite{2004MNRAS.352.1439H,2023RAA....23j4002W} & 260.133 & 2.859\\
  J1854+1050 & 207.2 & \cite{1986ApJ...311..694S,2024ApJS..271...23D} & 206.657 & 2.591\\
  J1859+00 & 420.0 & \cite{1996ApJ...469..819C,2023RAA....23j4002W} & 422.287 & 0.967\\
  J1901+0331 & 401.2 & \cite{1970Natur.227.1123D,2023RAA....23j4002W} & 400.983 & 0.308\\
  J1902+0556 & 177.486 & \cite{1974ApJ...191L..59H,2004MNRAS.353.1311H} & 177.065 & 1.309\\
  J1902+0615 & 502.9 & \cite{Lyne81,2023RAA....23j4002W} & 502.838 & 0.516\\
  J1903+0135 & 245.2 & \cite{1973NPhS..244...84D,2023RAA....23j4002W} & 244.842 & 0.071\\
  J1905+0709 & 245.3 & \cite{1986Natur.320...43C,2023RAA....23j4002W} & 245.112 & 2.89\\
  J1906+0641 & 472.8 & \cite{1986Natur.320...43C,2023RAA....23j4002W} & 472.032 & 2.647\\
  J1908+0500 & 201.4 & \cite{1995ApJ...449..156N,2023RAA....23j4002W} & 201.296 & 0.204\\
  J1909+0641 & 36.7 & \cite{2009ApJ...703.2259D,2023RAA....23j4002W} & 36.418 & 0.585\\
  J1909+1102 & 149.98 & \cite{1973NPhS..244...84D,2024ApJS..271...23D} & 149.739 & 0.355\\
  J1910+0358 & 82.93 & \cite{1975MNRAS.171P..17M,2004MNRAS.353.1311H} & 82.702 & 1.115\\
  J1913+0446 & 109.1 & \cite{2003MNRAS.342.1299K,2023RAA....23j4002W} & 109.565 & 0.136\\
  J1914+0219 & 233.8 & \cite{2006MNRAS.372..777L,2023RAA....23j4002W} & 234.151 & 2.443\\
  J1915+1009 & 241.693 & \cite{1975ApJ...201L..55H,2004MNRAS.353.1311H} & 241.363 & 1.257\\
  J1916+0951 & 60.9 & \cite{1974ApJ...191L..59H,2023RAA....23j4002W} & 60.948 & 0.758\\
  J1916+1312 & 237.0 & \cite{1974ApJ...191L..59H,2023RAA....23j4002W} & 236.891 & 0.3\\
  J1917+1353 & 94.5 & \cite{1971IAUC.2356R...1S,2023RAA....23j4002W} & 94.617 & 0.044\\
  J1918+1444 & 27.2 & \cite{1974ApJ...191L..59H,2023RAA....23j4002W} & 27.154 & 0.022\\
  J1919+0021 & 90.3 & \cite{1972Natur.240..229D,2023RAA....23j4002W} & 90.207 & 0.116\\
  J1921+1419 & 91.6 & \cite{1974ApJ...191L..59H,2023RAA....23j4002W} & 91.524 & 1.301\\
  J1921+1948 & 154.2 & \cite{1973NPhS..244...84D,2023RAA....23j4002W} & 153.806 & 2.319\\
  J1921+2153 & 12.4 & \cite{1968Natur.217..709H,2023RAA....23j4002W} & 12.462 & 0.102\\
  J1922+1733 & 234.0 & \cite{2013MNRAS.434..347L,2023RAA....23j4002W} & 233.982 & 0.132\\
  J1922+2110 & 217.0 & \cite{1973NPhS..244...84D,2023RAA....23j4002W} & 216.794 & 2.023\\
  J1923+4243 & 52.9 & \cite{2014ApJ...791...67S,2023RAA....23j4002W} & 53.131 & 0.298\\
  J1926+1648 & 176.8 & \cite{1974ApJ...191L..59H,2023RAA....23j4002W} & 176.898 & 0.417\\
  J1932+1059 & 3.18 & \cite{1968Natur.220..753L,2024ApJS..271...23D} & 3.189 & 0.007\\
  J1932+2020 & 211.1 & \cite{1974ApJ...191L..59H,2023RAA....23j4002W} & 211.342 & 2.351\\
  J1932+2220 & 219.2 & \cite{1975ApJ...201L..55H,2023RAA....23j4002W} & 218.871 & 0.684\\
  J1935+1616 & 158.5 & \cite{1970MNRAS.149..301D,2023RAA....23j4002W} & 158.578 & 0.048\\
  J1937+2544 & 53.2 & \cite{1985ApJ...294L..25D,2023RAA....23j4002W} & 53.247 & 0.097\\
  J1939+2449 & 142.88 & \cite{1986ApJ...311..694S,2024ApJS..271...23D} & 142.97 & 0.074\\
  J1943+0609 & 70.7 & \cite{2001MNRAS.326..358E,2023RAA....23j4002W} & 70.644 & 0.725\\
  J1946+1805 & 16.1 & \cite{1970Natur.225..167V,2023RAA....23j4002W} & 16.115 & 0.47\\
  J1946+2611 & 165.0 & \cite{1995ApJ...449..156N,2024ApJS..271...23D} & 165.02 & 2.538\\
  J1948+3540 & 129.37 & \cite{1970MNRAS.149..301D,2024ApJS..271...23D} & 129.042 & 0.148\\
  J2022+2854 & 24.63 & \cite{1973AA....27...67B,2024ApJS..271...23D} & 24.658 & 0.049\\
  J2150+5247 & 148.9 & \cite{1985ApJ...294L..25D,2023RAA....23j4002W} & 149.603 & 0.248\\
\hline\end{tabular}
\end{table*}

\begin{figure}[h]
    \centering
    \includegraphics[width=0.95\linewidth]{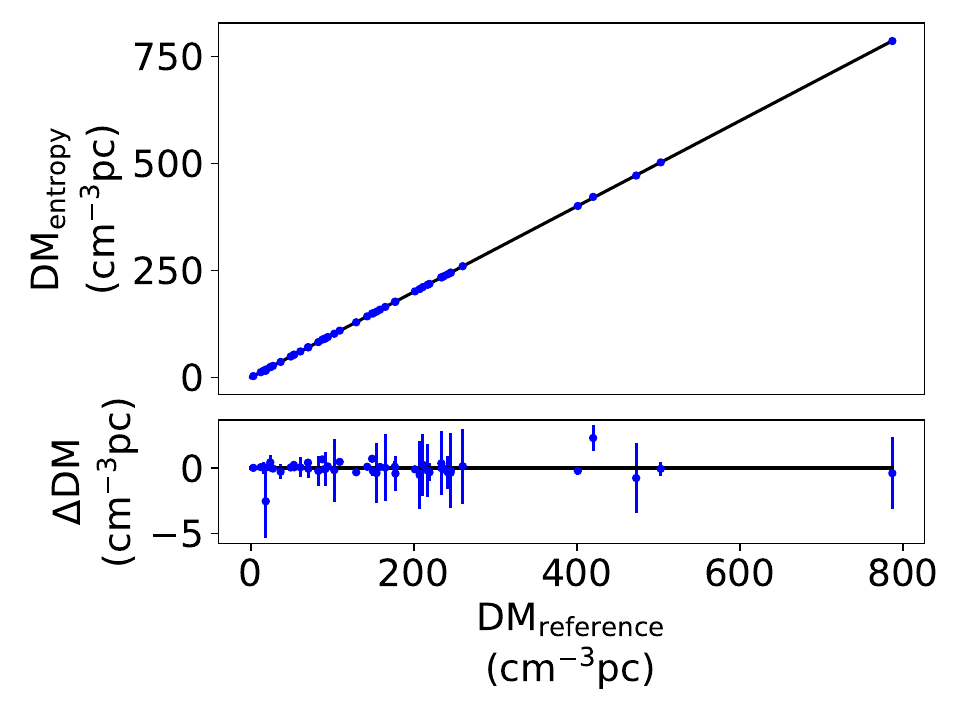}
    \caption{An comparison between the DMs from the latest references.}
    \label{compare}
\end{figure}

Figure~\ref{compare} compares the DMs from the latest references, compiled by \cite{2005AJ....129.1993M}\footnote{\url{https://www.atnf.csiro.au/people/pulsar/psrcat/}}, with those measured in this work. 
Among them, the DM values of 13 pulsars differ from those of the references within the error range, but the discrepancies for 10 of them are still less than 0.5\%. For 3 stars, the discrepancies are greater than 0.5\%, but their DMs instead align with those reported in the older references. These 3 stars are noted in Table~\ref{different}.

\begin{table*}[h]
    \caption{3 pulsar DMs align with older references}
    \label{different}
    \centering
    \begin{tabular}{|l|r|r|l|r|l|}
\hline
  \multicolumn{1}{|c|}{PSR} &
  \multicolumn{1}{c|}{Entropy DM} &
  \multicolumn{1}{c|}{Latest DM } &
  \multicolumn{1}{c|}{Latest Reference} &
  \multicolumn{1}{c|}{Old DM} &
  \multicolumn{1}{c|}{Old Reference} \\
\hline
  J0628+0909 & 88.557 & 87.9 & \cite{2023RAA....23j4002W} & 88.3 & \cite{2013ApJ...772...50N}\\
  J1851+1259 & 70.605 & 70.2 & \cite{2023RAA....23j4002W} & 70.6333 & \cite{2016AA...591A.134B}\\
  J1859+00 & 422.287 & 420.0 & \cite{2023RAA....23j4002W} & 423.0 & \cite{2020ApJ...892...76M}\\
\hline\end{tabular}
\end{table*}
\section{Disscusions and Conclusions}

The fitting method will make the profile tend to the model used, so the fitting method is not suitable when studying the single-pulse profile. Instead, the profile discrimination method should be used to judge the most real profile, and at the same time the optimal DM can be obtained. This work considers that information entropy can be used as an indicator to determine pulse profiles, i.e., we believed that the pulse profile with the minimum entropy, which contains the maximum amount of information, is closest to the real profile. This work uses real data from 48 pulsars to demonstrate the effectiveness of the minimum entropy criterion, but it does not mean that this criterion is optimal. If one wants to study pulse profiles, it is recommended to adopt multiple indicators for research, respectively. If the most suitable indicator can be found in the future, it will imply certain characteristics of the radio source radiation mechanism.

It needs to be clarified here regarding the discrepancy in DM for the same pulsar reported in different references. If one wishes to study pulsar profiles, it is advisable to calculate the DM for each pulse individually, with reference values serving merely as a pre-input parameter. If one aims to investigate the relationship between DM and the interstellar medium (ISM) along different line-of-sight directions or to study the temporal variation of DM for the same radio source, the contributions of the Earth's atmosphere and the Sun to DM should be considered. Not all references account for the contributions of the Earth's atmosphere and the Sun to DM, as this is quite complex and varies with time and location. Even if some reference does consider the effects of the Earth's atmosphere and the Sun, systematic errors may exist between different correction methods. Therefore, it is best not to directly use the DM values from the references when studying the relationship between the DM and the ISM but rather to reprocess all data using a consistent method. In addition, different data preprocessing methods, computational processes, and DM calculation methods can exert a certain influence on the final DM. Hence, it is not crucial that the DM provided in references differs, as long as the discrepancy is not too large to provide a good pre-input parameter.
\begin{acknowledgements}
   The authors appreciate helpful discussions with Prof. JinLin Han, Dr. Tao Wang, DeJiang Zhou, and Weicong Jing for data preparation.   
\end{acknowledgements}

\onecolumn
\begin{appendix}
\section{An example code}
\begin{verbatim}
import numpy as np
import astropy.io.fits as fits
import pandas as pd
import matplotlib.pyplot as plt
import os
import copy
from functools import partial
from multiprocessing import Pool,Manager
def func(binsize,y,dH,min_time,max_time,idm):
    print(idm,end='\r')
    idt=idm*(1/dH['mhz']**2-1/y[0]**2)/2.41E-4
    dH['newtime']=dH['oritime']-idt
    dH=dH[(dH['newtime']>min_time)&(dH['newtime']<max_time)]#make sure the data is a rectangle \
    then calculate entropy
    phase=binsize*0.1
    dH.loc[:, 'No'] = (dH['newtime'] - phase) // binsize#Subtracting a phase to prevent the bins \
    containing data points from being excessively sensitive to DM, which may lead to resonance \
    and form periodicity. This issue can be avoided by distributing each data point into multiple \
    bins based on certain principles. Improvements can be made if you think it is necessary.
    bindata=dH.groupby('No')['flux'].sum().reset_index()
    binflux=bindata['flux'].values
    binflux=binflux/np.sum(binflux)
    entropy=-np.sum(binflux*np.log2(binflux))
    return [entropy]
def error(dms,y):
    binsize=np.max(dms)-np.min(dms)
    binsize=binsize/20#It will affect the smoothness of the results, and consequently influence \
    the error range. However, as long as it remains unchanged, the final error obtained can be \
    reliable for weighting.
    sx=[]
    sy=[]
    for i in range(len(y)):
            l1=np.where((dms>=dms[i]-0.5*binsize)&(dms<dms[i]))
            l1=y[l1]
            l2=np.where((dms<=dms[i]+0.5*binsize)&(dms>dms[i]))
            l2=y[l2]
            if len(l1)<0.6*len(l2):
                    sx.append(dms[i])
                    sy.append(np.nan)
                    continue
            if len(l2)<0.6*len(l1):
                    sx.append(dms[i])
                    sy.append(np.nan)
                    continue
            l=l1.tolist()+l2.tolist()+[y[i]]
            sx.append(dms[i])
            sy.append(np.median(l))
    cha=y-sy
    cha=cha[cha==cha]
    std=np.std(cha)
    out=np.where(y-np.min(y)<=3*std)
    plt.title(str(np.round(dms[y==np.min(y)][0],2))+'('+str(np.round(np.min(dms[out]),2))+', ' \
    +str(np.round(np.max(dms[out]),2))+')')
    plt.plot(sx,sy)
    plt.plot(dms,y)
    plt.show()#show entropy-DM curve
    return dms[y==np.min(y)][0],np.min(dms[out]),np.max(dms[out])
if __name__=='__main__':#Required for parallel operation
    min_dm=#Lower limit of the DM. If you do not know, set it to 1.
    max_dm=#Upper limit of DM. If you do not know, set it to a large number, but if it is too \
    large, you can only analyze a small time length of the waterfall.
    h=#step of DM. The smaller the step size, the higher the accuracy, and the larger the step \
    size, the faster the calculation.
    y=#frequence list. unit: mhz. y[0] is the lowest freq and y[-1] is the highest freq.
    t=#time step of the data. unit: second.
    waterfall=#Two dimensional matrix corresponding to time and frequency. if you plt.imshow(H), \
    the x-axis is frequency, and the y-axis is time.
    max_time=len(waterfall)*t
    min_time=- (max_dm *(1/y[-1]**2-1/y[0]**2)/2.41E-4)
    if min_time>=max_time:
        print('Your max_dm is too large! Your data need a longer observation time. Now your \
        max_dm is '+str(np.abs(max_time/((1/y[-1]**2-1/y[0]**2)/2.41E-4))))
    dms=np.arange(min_dm,max_dm,h)
    allH=pd.DataFrame()#prepare to transform matrix to data points.
    pool=Pool()#Initialize parallel calculation. If you do not like parallel, \
    pool=Pool(processes=1)
    data=waterfall#If you want to remove interference, add your function here.
    me=np.median(data)
    sig=np.std(data/me)
    x=range(len(data))
    x=np.array(x)
    nx,ny=len(x),len(y)
    ally=np.empty(len(x)*len(y))
    allf=np.empty(len(x)*len(y))
    allorix=np.empty(len(x)*len(y))
    indices=np.arange(nx*ny)
    ddt=y-y
    orix=x[:,np.newaxis]*t-ddt[np.newaxis,:]
    allorix[:]=orix.T.flatten()[indices]
    ally[:]=np.repeat(y,nx)[indices]
    allf[:]=data.T.flatten()[indices]
    allH['oritime']=allorix
    allH['mhz']=ally
    allH['flux']=allf
    allH['snr']=allH['flux']/me
    allH=allH[allH['snr']>1+3*sig]#In this code we only analyze data points larger \
    than 3*sigma, so a DataFrame is used to save the points.
    binsize=t#Calculate the profile from a waterfall, you can change it, but \
    binsize=t is ok.
    dmfunc=partial(func,binsize,y,allH,min_time,max_time)
    result=pool.map(dmfunc,dms)
    result=np.array(result)
    result=result.T
    entropy=result[0]
    mid,low,high=error(dms,entropy)#calculate the error range. You can refine function \
    error() as you wish.
    print((mid,low,high))
    pool.close()
    pool.join()
    print('Done!')
\end{verbatim}
\end{appendix}
\twocolumn




\onecolumn

\FloatBarrier 
\clearpage

\end{document}